\documentclass[12 pt]{article}
\usepackage{times}
\usepackage{subfigure}

\usepackage{pifont}
\usepackage{floatrow}
\usepackage{caption}
\usepackage{mathrsfs}
\usepackage[fleqn]{amsmath}
\usepackage{amsfonts,amsthm,amssymb,mathrsfs,bbding}
\usepackage{txfonts}
\usepackage{graphics,multicol}
\usepackage{graphicx}
\usepackage{color}
\usepackage{caption}
\usepackage{indentfirst}
\usepackage{cite}
\usepackage{latexsym,bm}
\usepackage{enumerate}
\usepackage{epsfig}
\evensidemargin=\oddsidemargin\topmargin=-1.5cm

\theoremstyle{definition}

\newtheorem{remark}{Remark}

\addtocounter{section}{0}

\numberwithin{equation}{section}

\let\trueiiint=\iiint
\def\iiint{\mathop{\textstyle\trueiiint}\limits}
\def\intinfty{\int\limits_{\!\!-\infty\,\,}^{\,\,\infty\!\!}\kern-0.0em}
\def\iintinfty{\mathop{\int\!\!\int}\limits_{\!\!-\infty\,\,}^{\,\,\infty\!\!}\kern-0.0em}
\def\iiintinfty{\mathop{\int\!\!\int\!\!\int}\limits_{\!\!-\infty\,\,}^{\,\,\infty\!\!}\kern-0.0em}
\def\Real{{\mathbb R}}

\let\<=\langle
\let\>=\rangle
\let\^=\hat
\def\~#1{{\-ox{\sf#1}}}
\def\N{{\mathbb N}}

\let\-=\mathbf

\def\@#1{{\cal #1}}

\def\N{{\mathcal{N}}}
\def\Cov{\mathrm{Cov}}

\let\<=\langle
\let\>=\rangle

\let\^=\hat

\def\circ{\ifmmode\mathchar"220E\else$\mathchar"220E$\fi}
\def\@#1{{\cal #1}}
\def\x{{\bm{x}}}
\def\y{{\bm{y}}}
\def\u{{\bm{u}}}
\def\r{\bm{r}}

\newcommand{\rev}[1]{{\color{black}{#1}}}

\begin{document}
\title{Explicit Estimation of Derivatives from Data and Differential Equations by Gaussian Process Regression}

\author{Hongqiao Wang$^{\ddag}$
\quad {and} \quad
Xiang Zhou$^{~\S}$ \\[2mm]
{\it $^{\ddag}$School of Mathematics and Statistics} \\
{\it Central South University} \\
{\it Changsha 410083, P.R. China} \\
[2mm]
{\it $^{~\S}$School of Data Science and Department of Mathematics}\\
{\it  City University of Hong Kong}\\
{\it Kowloon, Hong Kong SAR} 
}


\date{}

\maketitle {\flushleft\large\bf Abstract }

In this work, we employ the Bayesian inference framework to    
robustly estimate  the derivatives of a function from       
    noisy  observations of only the function values  at given location points,  
under the assumption of a  physical model  in the form of   differential equation
governing the function and its derivatives.
To overcome the instability of numerical differentiation  
of the  fitted function solely from the data   or  the prohibitive   costs of   
solving the differential equation on the whole domain,  we use the Gaussian processes to jointly model the solution, the derivatives, and the differential equation, by utilising the fact that differentiation is a linear operator.
 By regarding the linear differential equation as a linear constraint, 
we develop the Gaussian process regression with constraint method (GPRC)  at Bayesian perspective to improve the prediction accuracy
 of   derivatives. 
    For nonlinear  equations,
    we propose a Picard-iteration  approximation of linearization around  the Gaussian process
    obtained only from data to iteratively apply our GPRC.
    Besides, 
    a product of experts method is applied   if  the initial or boundary condition  is also available. We present several numerical results   to illustrate the  advantages   of our new method   and show 
  the  new 
    estimation of the derivatives from GPRC 
    improves  the parameter identification  with
    less data samples.

\begin{flushleft}
\end{flushleft}

\section{Introduction}\label{s:intro} 

 \rev{ To infer a nonlinear function from  its noisy measurements at a given set of inputs is a classical     statistical learning problem and  
a vast of well-established methods, ranging from  polynomial and spline to kernel method and   neural network, are available for many important applications\cite{hastie2009elements}.
 However, if the interest  is the derivatives and there is no additional observations of derivatives,
the problem of estimating   derivatives is more challenging and subtle than  estimating the function itself and less attentions have been  given  to this issue.  When the measurements are  on the gridded  data,  the naive method of finite difference deteriorates the accuracy of  numerical   derivatives and the algorithm may become extremely unstable if the data is scarce  and the order of derivatives is high. 
The modern machine learning approach \cite{Vapnik1998,hastie2009elements}  enforces certain regularity in learning the function
and practically relies on 
cross-validation technique to select the right model.  But the lack of data from derivatives generally hinders the  direct use of this framework.
 In applications, many practitioners simply take the symbolic or numerical differentiation  after the function is trained first, and seem to
pay less attention to the accuracy and robustness of 
the obtained derivatives.
} 

\rev{
As we will explain soon,  many applications in need of  robust estimation of derivatives  arise from  various problems related to    differential equation (DE) models  which 
the function and its derivatives should satisfy.  This situation in fact offers  an extra advantage for estimating the function and derivatives.
In this paper, we shall assume that the observation data are the measurements (subject to uncertainty noise) of the solution of the DE model  and  the 
(partial or complete) information of the underlying DE model is also known to us.
We study the problem of efficiently and  robustly estimating  the function and its derivatives up to the order of interest within a Bayesian inference framework,
by taking into account of the solution observations and the corresponding DE model   simultaneously and intrinsically.
In the next, before we present our main ideas,  we provides some further  backgrounds on the importance of estimate of derivatives.
}

  Differential equations which include ordinary differential equation (ODE) and partial differential equation (PDE) are used to   model a wide variety of physical phenomena  such as heat transfer, electromagnetism, and structural deformations. 
In physics or chemistry problems, the state, which is   the solution of the model, and its derivatives (time derivative or spatial  gradient)   usually have 
specific meanings, e.g.,  location and velocity in Newton mechanics,  electric potential and electric field in electrostatics.
In ODE models, often the time derivative of a state variable is of as much interest as the state variable itself \cite{swain2016inferring}.
 Many physical constitutive laws  appear  in form of PDEs 
   linking  the    derivatives to the state variable.
In the deterministic PDE, the derivatives are generally   computed by numerical differentiation techniques after  the  solution is   computed  
from numerical methods like  finite element method, for instance.
 In general,  the  convergence order of   errors for such derivatives  is   one order lower than the convergence rate for the   solution itself.
 In history, this accuracy degradation  problem for the gradient is 
 alleviated to certain extent  by  the mixed finite element method  (\cite{MixFEM_Brezzi_Survey}), which treats the gradient   as an independent   variable and constructed  an extended   system jointly for the solution and its gradient.
  When the PDE models  have random or missing parameters/coefficients, 
the estimate of the solution and the derivatives of the solution becomes a non-trivial task
because the numerical solution is typically a random function with uncertainties inherited  from 
the randomness of the PDE or the noisy measurement. 
Various numerical experiments 
suggest that 
a direct numerical difference scheme  for the derivative for a random function 
usually lacks of robustness and the higher order the derivatives, 
the less accuracy of the estimation. 

\rev{The demand  of robust estimation of derivatives becomes more critical 
when a DE model is supplemented with   available data
as the measurement of the state variable, while the DE model
is too complicated to solve  precisely or  is  
incomplete with missing parameters or with random coefficients.  One  of such scenarios    is 
the problem of  identifying  the missing information of  a DE model
from  the  noisy measurement of the state variable} \cite{varah1982spline,Ramsay2007,liang2008parameter,xun2013parameter,brunton2016,rudy2017data,raissi2018hidden,raissi2017machine}.
 In these problems, the noisy measurements of the solution   are given at randomly drawn or deliberately chosen points
 and    optimization frameworks are proposed   to identify  the missing information in the DE.   In these optimization problems,
  the state variables {\it and} its all derivatives appearing in the DEs
  are required as the input information, 
however  there is {\it no measurement data for the derivatives}.
  \rev{One mainstream approach is  the two-stage procedure}    \cite{varah1982spline,liang2008parameter,poyton2006parameter}, which in the first stage estimates the function and its derivatives from noisy observations using data smoothing methods without considering differential equation models, and then in the second stage identifies   the missing parameters   by the method of least squares.
Many data smoothing methods are applicable in the first stage to fit the state function, such as  polynomial interpolation \cite{rudy2017data}, local polynomial regression \cite{liang2008parameter} and spline-based approach \cite{varah1982spline}.
 These methods are easy to implement and can achieve excellent  fitting  result for the state function itself, however the  
 derivatives of the fitted function in fact are not as accurate and robust as the function itself, which may significantly 
 affect the parameter identification in the second stage.
 So,     the main drawback of the two-stage method  is that the underlying differential equation is not  fully  utilized.
 
\rev{ There have been   efforts to  incorporate  the DE model in the first stage   to improve the accuracy of both the state function and its derivatives appearing in the DE model.}
In \cite{xun2013parameter} X. Xun {\it et al}   proposed to  incorporate the differential equation as a penalty term  in the first stage,
where  the solution is expressed  by B-spline basis functions
and the derivatives are based on   analytical derivatives of B-splines.
This produces more reliable derivatives for the second stage and finally improves the accuracy of estimating parameters significantly, particularly for the parameters associated with derivatives.
   However, this approach of treating differential equation as  a squared penalty gives
  rise to  difficult  optimization  problem or   Markov chain Monte Carlo  step  \cite{xun2013parameter}.

\rev{In view of the critical importance of robust estimation of derivatives for the above applications, 
we focus on how to combine the data of the state function and the DE information to 
improve the derivative estimation by a new approach.}
We propose a new statistical approach based on Gaussian process regression (GPR) to 
efficiently and robustly predict the derivatives in need. 
Compared with the basis-function expansion method and MCMC Bayesian approach, the Gaussian process (GP) model  \cite{seeger2004gaussian} is  flexible and efficient     for fitting noisy data.  The prior knowledge can also be easily encoded by the covariance of the GP.
\rev{
More importantly, the derivatives of a GP is also a GP since taking derivatives  is a {\it linear} operation.
When the DE model is also linear, we can gain   favourable	advantages from  this linearity property  of GP.
}

\rev{To achieve our goal,  we build our  method, with the name {\it{Gaussian Process Regression with Constraint}} (GPRC),  on the following ideas.} Firstly, we treat the state variable and the derivatives appearing in the differential equation
 jointly as a multi-dimensional  GP   to leverage  the advantage of  a jointly Gaussian distribution over the function and its derivatives
 \cite{seeger2004gaussian}.  Secondly,  to take  account  of  the differential equation that 
 state and derivatives should satisfy,  \rev{we exploit the fact that the residual of the differential operator corresponding to the DE should be zero,
 which provides  the zero-valued measurements of the residual. We    model  the residual for linear DEs
as  a mean-zero prior Gaussian  with a small   variance. Then, the observations for training the GP are
the paired values of the state variable and the residual.
  In other words, the available information of DE is treated as additional constraint for the multi-dimensional  GP of state and derivatives should satisfy.
Thirdly,  to generalize  our method to  nonlinear equations, we consider a Picard approximation of linearization
in which the nonlinear part  is approximated by linearization around a function obtained from the standard GPR first without using the equation,
and the multi-dimensional  GP is then recursively updated based on the linearized equation.
Lastly, in the prediction for the function and its derivatives on a new test location, 
the zero-valued residual observation from the DE model on the test location is also exploited to 
further strengthen the prediction accuracy of the derivatives.}
In addition, if  the linear boundary/initial  condition (BC/IC) is available, 
they can also be added as one of  linear constraints that the  GP should satisfy like the DE itself, by 
the product of experts method \cite{mayraz2001recognizing,Calderhead2009,barber2014gaussian}.
 Our method can be applied to the   parameter identification  problems but 
it certainly fits in  any application which needs a robust scheme of  derivative estimation
when a differential equation model underlying the data
 is available.

\rev{In our numerical experiments, we shall demonstrate the effectiveness of this new method  GPRC
by comparing the obtained derivatives with  
the results from the traditional GPR on a few linear DE models.  
We also present  a nonlinear  Van der Pol equation with a parameter $\mu$. Based on the  dataset  corresponding to the 
ground-truth $\mu^*=0.5$, we not only show a better estimation of the derivatives for this $\mu^{*}$, but also  illustrate
the supremacy  of our method when applied  to identify $\mu^*$ 
when the observations are scarce.
In this example,  the traditional GPR fails to find the correct 
$\mu^*$ because of the inaccurate estimation of derivatives. }

In the last part of this section, we selectively comment on  some  general  works related to the GP.  
For the use of  GP to solve  differential equations, 
there has been a considerable amount of literature. For example,   \cite{theore2003Solving} has used  GP to solve  linear differential equations  with noisy forcing terms.
The use of GPR for inference related to differential equation 
is intensively studied in the communities of machine learning 
\cite{sarkka2011linear, barber2014gaussian} and statistics  \cite{liang2008parameter,xun2013parameter}.
For the  idea of exploring the GP for the state function and the state-derivatives jointly
rather separately,
   \cite{Calderhead2009} focused on a Bayesian inference  of the parameters in  ODE models.
   \cite{Gao2008}  used the GP prior in Bayesian method to infer  parameters   in the ODE models of chemical networks. 
   For recent advancement of parameter identification, 
    \cite{brunton2016,rudy2017data}  worked on the   PDE ``discovery'' problem by 
   $\ell_0$ or $\ell_1$ optimization  to learn a spars set  of  coefficients in   PDE.  
  \cite{raissi2017machine,raissi2018hidden} 
worked  with the scarce and noisy measurement of both the state variable and a black-box forcing term  in the PDE
 and the joint GPR model  is applied to solve the solution function and infer the forcing-term function.
 But in existing  literature, we have not yet seen a
 specific method for efficiently and robustly  processing the derivative estimation
if the differential equation is also given.

The structure of the paper is organized as follows: 
We first formulate the problem setup in Section \ref{set:Problemsetup}. Our GPRC method for estimating the state and its derivatives is presented in Section 
\ref{set:method}. A product of experts method for initial or boundary conditions is introduced in Section \ref{set:IBC}. Numerical examples are presented in Section \ref{sec:examples} to demonstrate the effectiveness of the proposed method and the   extension to    nonlinear differential equation problems (Section \ref{VanderPol}).
Lastly, in  Section \ref{ssec:ip}, we use the nonlinear toy model  to  illustrate 
    the new estimation of derivatives by GPRC 
   improves the quality the parameter identification with less data samples.
  Section \ref{sec:conclusions} offers some closing remarks and outlooks.

\section{Problem setup}
\label{set:Problemsetup}

In general we formulate the differential equation system as a multidimensional dynamic process and view the ordinary differential equation as a one-dimensional  partial differential equation case. The solution function  is denoted as $u(\x)$, where $\x = (x_1, \dots,x_D)^T\in \mathbb{R}^D$. The PDE is modeled as 
\begin{equation}
\label{eq:pde}
\mathcal{F}(\x, u, \frac{\partial u}{\partial x_1},\dots, \frac{\partial u}{\partial x_i},\dots, \frac{\partial^2 u} {\partial x_i\partial x_j},\dots) = 0,
\end{equation}
 where the left-hand side of Eq. \eqref{eq:pde} consists of the state $u(\x)$ and its partial derivative terms ($\frac{\partial u} {\partial x_i}, \frac{\partial^2 u} {\partial x_i\partial x_j}, \dots, $). 
In practice, we only can directly observe the noisy observation  $y(\x)$ instead the state $u(\x)$ . We assume that $u(\x)$ is observed with measurement error and specifically, the noisy measurements satisfy 
\begin{equation} 
\label{y}
y_i = u(\x_i) + \epsilon_i,
\end{equation}
where $i = 1,\dots, n$, is the observation index, each point $\x_i\in \Real^D$ and $\epsilon_i$ is the measurement error, also called noisy/observation error. Assumption is made that the error $\epsilon_i$ is an  {\it iid} random variable and follows a Gaussian distribution with mean $0$ and variance $\sigma_u^2$.

Our objective is to   estimate the state $\bm{u}$ and its derivative terms $\frac{\partial^K \bm{u}}{\partial^k x_i \partial^{K-k}x_j}$, $k=0,\dots,K$, $K=1,2,\dots$, from the noisy data $(\x_i, y(\x_i))$ together with the equation \eqref{eq:pde},  and to quantify the uncertainty of our estimations.
The estimation problem in linear dynamic process where $\mathcal{F}$ in  \eqref{eq:pde} is a linear operator
 is more   fundamental and has been serving as the very start of nonlinear equations\cite{raissi2017machine,sarkka2011linear,theore2003Solving}.
 To discuss the linear equation case first, 
 we use  $\mathcal{L}u(\x)$ to denote    $\mathcal{F}(\x, u, \frac{\partial \u}{\partial x_i},\dots, \frac{\partial^2\u} {\partial x_i\partial x_j},\dots) $  as a function of $\x$
 if $\mathcal{F}$ is a linear operator.
\section{Methodology}
\label{set:method}
The proposed   algorithm for estimating the state and its derivatives, employs Gaussian process prior that is tailored to the corresponding differential operators.
\subsection{Gaussian process regression}
Specifically, the method starts by  assuming  that $u(\x)$ is a  Gaussian process  with
the zero mean    and  the covariance function $k_{uu}(\x,\x';\bm{\gamma})$, which is denoted as 
\begin{equation}
\label{eq:u_gp}
u(\x)\sim \mathcal{GP}(0, k_{uu}(\x,\x';\bm{\gamma})),
\end{equation}
where $\bm{\gamma}$ denotes the hyper-parameters of the kernel function $k_{uu}$. 
This means that $\mathbb{E}[u(\x)]\equiv 0$ and {$\Cov(u(\x),u(\x'))=k_{uu}(\x,\x';\bm{\gamma})$}.
The kernel $k_{uu}$ allows us to encode any prior knowledge we may have about $u(\x)$, and can accommodate the approximation of arbitrarily complex functions. 
The choice of the specific form of  $k_{uu}$ will be discussed later.

The key property of Gaussian process in our favor is that any linear transformation, such as differentiation and integration, of a Gaussian process  is still a Gaussian process. 
With the assumption \eqref{eq:u_gp},  we consider the
linear differential operator, $ \mathcal{L} $, acted on  $u(\x)$.
Then the function $\mathcal{L} u(\x)$ is also a mean-zero Gaussian process 
\begin{equation}
\mathcal{L} u(\x) \sim \mathcal{GP}(0, k_{LL}(\x,\x')) 
\end{equation}
where {$k_{LL}(\x,\x')=\Cov(\mathcal{L}u(\x), \mathcal{L}u(\x'))$} denotes the covariance function  of   $\mathcal{L}  u$ between    $\x$ and    $\x'$.
 The following fundamental relationship between the kernels $k_{uu}$ and $k_{LL}$ is well-known
(see  e.g. \cite{seeger2004gaussian,theore2003Solving}),
\begin{equation}
\label{eq:k_LL}
k_{LL}(\x,\x';\bm{\gamma}) = \mathcal{L}_\x \mathcal{L}_{\x'} k_{uu}(\x,\x';\bm{\gamma}).
\end{equation}
Here  we add the subindex in  the linear differential operator $ \mathcal{L} $  to specify  the differentiation is for $\x$ or ${\x}'$ variable
in the kernel function.
$k_{uu}$ and $k_{LL}$ share the same hyper-parameter $\bm {\gamma}$.
Simiarly, 
for the covariance between   $u$ and   $\mathcal{L} u$,
 $k_{uL}(\x,\x')=\Cov(u(\x), \mathcal{L}u(\x'))
$ and $k_{Lu}(\x,\x')  = \Cov(\mathcal{L}u(\x), u(\x'))$, we have 
\begin{equation}
k_{uL}(\x,\x';\bm{\gamma})= \mathcal{L}_{\x'}k_{uu}(\x,\x';\bm{\gamma}),
~\text{ and } ~
k_{Lu}(\x,\x';\bm{\gamma}) = \mathcal{L}_{\x}k_{uu}(\x,\x';\bm{\gamma}).
\end{equation}

 Since we shall work on the differential equation $ \mathcal{L}u (\x)=0$,
 we  introduce a random function  $r(\x)$ as the residual 
of the linear differential equation $ \mathcal{L}u=0$ for convenience: 
\begin{equation} \label{r}
r (\x) := \mathcal{L}u(\x).
\end{equation}
We refer the original equation as the {\it{equation constraint}},  $r(\x)=0$, and the equation  will be interpreted later as the observation of zero values of the function $r(\x)$ at any point $\x$, in a similar way to the observation $y_i$ of $u(\x)$
at $\x_i$.  
With the given prior of the GP $u(\x)$, 
we then have the the  prior for the pair
$( u(\x), r(\x))$. The covariances above for $u$ and $\mathcal{L}u$ can be rewritten as 
\begin{align}
\label{eq:cov_ru}
\begin{split}
k_{rr}(\x,\x';\bm{\gamma}) &= \mathcal{L}_\x \mathcal{L}_{\x'} k_{uu}(\x,\x';\bm{\gamma}),\\
k_{ur}(\x,\x';\bm{\gamma})& = \mathcal{L}_{\x'}k_{uu}(\x,\x';\bm{\gamma}),\\
k_{ru}(\x,\x';\bm{\gamma}) &= \mathcal{L}_{\x}k_{uu}(\x,\x';\bm{\gamma}),
\end{split}
\end{align}
respectively. 

 So the equation residual $r(\x)$ is also a Gaussian process, whose kernel 
 is related to the derivatives of the kernel of  $u(\x)$. Based on the Gaussian assumption and the covariance expression between $u$ and $r$ in \eqref{eq:cov_ru}, a joint inference framework  of Gaussian process regression for the available observation  data of $u$ and $r$   can be naturally constructed.
 By interpreting  the equation $r=\mathcal{L}u=0$ as the constraint of the function $r$,
 we refer this approach as   {\it{Gaussian Process Regression with Constraint}} (GPRC). GPRC will significantly improve the accuracy of estimation of solution and its derivatives due to the additional equation information. The  advantages in  the comparison with standard Gaussian process regression (without constraint) will be shown in Section \ref{sec:examples}.
 
 \begin{remark}
 The linear differential operator   can be easily generalized to an {\it affine} operator,
 i.e., the equation $\mathcal{L}u(\x) =f (\x)$. Then the linear constraint $r(\x)=0$
 should be modified as  $r(\x)=f(\x)$. To be concise, we  present our ideas and methods 
 only for the linear case $f(\x)=0$. 
 \end{remark}

 The differential equation discussed above  is the linear constraint, i.e., $r=\mathcal{L}u$ is a linear mapping of $u$.  For the non-linear differential equation $\mathcal{F}(u)=0$, we propose a linearizaiton strategy to make GPRC applicable to nonlinear problems. 
To illustrate idea, consider  a special case
 $\mathcal{F}(u) = \mathcal{L}(u) + \mathcal{N}(u)$ for example,  where $\mathcal{N}$ is the nonlinear part.
 We apply the standard GPR only from the data $\{\x_i, y_i\}$ to train a GP $u_0$, and then 
 apply the GPRC to the affine constraint   $\mathcal{F}_0(u):=\mathcal{L}(u) + \mathcal{N}(u_0)=0$
 and   the data $\{\x_i, y_i\}$ to train $u$ as an update of $u_0$. This  approach can be implemented recursively and is  a type of  Picard iteration
 {\it per se}. The stopping criterion is
   based on the Root Mean Square Error (RMSE) of the true (nonlinear) residual. 
In  Section \ref{VanderPol}, we shall specifically show how to apply this idea to  a nonlinear oscillator equation. 

\subsection{Kernel}
\label{subsec:kernel}
 The kernel (covariance function) is the crucial ingredient in a Gaussian process predictor, as it encodes our assumptions about the function   we wish to learn. Without loss of generality, the Gaussian prior of the solution used in this work is assumed to have a squared exponential covariance function (other kinds of kernels are also suitable in this framework), i.e.,
\begin{equation}\label{37}
k_{uu}(\x, \x',\bm{\gamma}) = \gamma^2_\alpha \exp(-\frac{1}{2} ||\x-\x'||^2_{\bm{\gamma}_l})
\end{equation}
where $\gamma^2_\alpha$ is a variance parameter, $\x$ is a $D$-dimensional vector that includes spatial and/or temporal coordinates, the norm $||\cdot||_{\bm{\gamma}_l}$ is defined as 
\begin{equation}
\label{gam}
||\bm{v}||_{\bm{\gamma}_l} = (\bm{v}^T \bm{\gamma}_l\bm{v})^{\frac{1}{2}}, ~\bm{\gamma}_l = \mbox{diag}(\gamma_{l1}, \dots, \gamma_{lD}).
\end{equation}
 $\bm{\gamma}_l$ is the length scale parameter and $\bm{\gamma} = (\gamma_\alpha, \bm{\gamma}_l)$. The squared exponential covariance function chosen above implies smooth approximations. More complex function class can be accommodated by appropriately choosing kernels. For example, non-stationary kernels employing nonlinear warpings of the input space can be constructed to capture discontinuous response. In general, the choice of kernels is crucial and in many cases still remains an art that relies on one's ability to encode any prior information (such as known symmetries, invariant, etc.) into the regression scheme. In our problem here  related to the differential operator $\mathcal{L}$,
 we require that the kernel satisfies the regularity such that the derivatives of the kernel,
 $\mathcal{L}_\x\mathcal{L}_{\x'}k_{uu}(\x,\x')$,  which is the covariance function  of $\mathcal{L}u$,  are at least continuous. 
 Our choice of squared exponential covariance function 
 surely satisfies this requirement. 
 
The kernel $k_{Lu}(\x,\x';\bm \gamma)$ can be easily computed based on \eqref{37}. For instance the kernel of first order derivative term $\partial_\x u(\x)$  can be expressed as:
  \begin{equation}
  \label{klu}
  k_{Lu}(\x,\x';\bm \gamma):=\frac{\partial k_{u,u}(\x,\x';
  \bm \gamma)}{\partial \x} \\
 = -k_{uu}(\x,\x',\bm{\gamma}) \bm{\gamma}_l  (\x-\x') 
 \end{equation}
The expressions of other  kernel for a general linear differential operator, such as $\frac{\partial^2 k_{u,u}(\x,\x';\bm \gamma)}{{\partial\x}^2}$, $\frac{\partial^2 k_{u,u}(\x,\x';\bm \gamma)}{\partial \x \partial \x'}$  can be computed similarly.

Due to  the irreducible  measure noise
  in \eqref{y}, the covariance matrix in the prior of $u(\x)$, 
  $k_{uu}$ usually needs to be added with a noise kernel $\sigma_u^2\bm{I}_u$, $\bm{I}_u$ is identity matrix and the  
  parameter of variance   $\sigma_u^2$ can be optimized with the kernel parameters $\bm{\gamma}$ together. 
  In a similar style, we can also introduce a small parameter $\sigma_r^2 \bm{I}_r$
  for the residual function $r(\x)$.
\subsection{Training}
\label{training}

The training process is to   find the optimal parameters $\bm \gamma$ and $\sigma_u^2$ by maximum likelihood estimation (MLE). 
Since  all GPs  appearing  above share the same set  of hyper-parameters $\bm \gamma$,
this helps  reduce   computational burden.

Given $n$  noisy observations of the state $u$  
at  $n$ points  $\{\x_i: 1\leq i\leq n\}$,  we  denote  the state vector $\y \equiv [y_1, y_2,\dots,y_n]^T\in \Real^{n\times 1}$, the residual vector $\r \equiv [r_1, r_2,\dots,r_n]^T\in \Real^{n\times 1}$ and 
the training point matrix $X \equiv [\x_1, \x_2,\dots,\x_n]^T\in \Real^{n\times D}$.
Let   $Y = \begin{bmatrix} \y\\ \r\end{bmatrix} \in \Real^{2n}$. 
Then
 the   negative log marginal likelihood
of  $p(Y\vert \bm\gamma, \sigma^2_u)$ 
 has the following expression
\begin{equation}
\label{eq:logmarginal}
\begin{split}
-\log p(Y| \bm{\gamma}, \sigma^2_u) = \frac{1}{2}\log (\det{K})+\frac{1}{2}Y^T K^{-1}Y + \frac{n}{2}\log 2\pi,
\end{split}
\end{equation}
where  the $2n \times 2n$ matrix $K$ is defined by
\begin{equation*}
K = \begin{bmatrix}
&K_{uu}+ \sigma^2_u I & K_{ur}\\
&K_{ru} & K_{rr} + \sigma^2_r I
\end{bmatrix},
\end{equation*}
 The matrices $K_{uu}, K_{ur}$,  $K_{ru}$ and $K_{rr}$
 correspond to, respectively, the kernel functions $k_{uu}, k_{ur}, k_{ru}$ and $k_{rr}$ in \eqref{37} and \eqref{eq:cov_ru}, evaluated at the $n$ points $ \{\x_i\} $.  

To compute the optimal kernel hyperparameters $\bm{\gamma}$ and observation noise variance $\sigma^2_u$, a Quasi-Newton optimizer L-BFGS is employed to minimize the negative log marginal likelihood \eqref{eq:logmarginal}. Cholesky factorarization of $K$ is used to compute both the inverse and the determinant.

It is noteworthy that the constraint from the DE model should be rigorous, which  means $\sigma^2_r$ should  be exactly zero. However, for numerical stability
and statistical generalization, we add a regularization $\sigma^2_r$ term onto the constraint covariance matrix $K_{rr}$ and  then 
 $\sigma^2_r $  is   a hyper-parameter to tune.
 In principle,   the optimal  value of $\sigma^2_r$ is related to the complexity of model
 (e.g., the choice of the kernel),  the variance of the measurement noise (i.e., $\sigma_u^2$) in the data,
 and the DE model itself (e.g.,  the order of derivatives in appearance). 
  The practical optimal $\sigma^2_r$  can be determined by any standard 
 parameter-tuning method like cross-validation.
 A note is that $\sigma^2_r$ should be larger than $\sigma_u^2$ since
 taking  derivatives in general amplifies  the noise (with high frequency).
We  empirically find  a tenfold size of $\sigma_u^2$ seems satisfactory 
if the DE model has up to a second order derivative.

\subsection{Prediction}
\label{ssec:predict}
After the hyper-parameters in the GPRC are computed,
the prediction of the function $u(\x)$ or its derivatives of interest,
denoted as a function $l(\x)$ (e.g., $l(\x)=\partial_{\x} u(\x)$),
at a new  test point $\x_*$ is described  below.
The covariance function of the GP $l(\x)$ is denoted by $k_{ll}(\x,\x')$ by the convenction.

With a given   point $\x_*$,  the differential equation provides  the fact $r(\x_*)=0$, which
is a useful observation to incorporate the Bayesian inference.
To enhance this condition locally, we actually consider  a small neighbourhood  around  $\x_*$
and choose  $m$ \rev{equally-spaced} points in this neighbourhood.
   \rev{ The set $\chi$ is similar to the window/scale  concept in local polynomial regression and 
  thus adaptive strategy is possible  (\cite{GVK19282144X}).
We refer the collection of these $m$ points as the extended set  $\chi:=\{\x_*^j: 1\leq j\leq m\} $
and  by the differential equation, we have   the equation constraints   in this  extend   set, i.e., $\r_{\chi}=\bm 0$. 
 These points $\x_*^j$ are supposed to resolve a characteristic length for the residual process $r(\x)$. 
So,   the size of  the extend set $\chi$ and the number $m$  are related to the
 correlation length of the residual GP, specifically,  $1/\bm{\gamma}_l$ with $/\bm{\gamma}_l$
 and their choice is a balance between computational cost and the prediction accuracy from the DE model.
 In practice, just a few number of $m $ in each dimension is sufficient and we will give some details in Section 
  \ref{sec:examples}.
}

Let  $u_*$ and $l_*$ represent the state and its derivative at the test point $\x_*$, respectively. Adding the equation constraint vector $\r_{\chi}$ at the points in the extend set $\chi$,  we have the joint distribution for the Gaussian priors
\begin{equation}
\label{eq:joint_eq}
\begin{bmatrix} \begin{bmatrix}\y \\ \r_{\chi} \end{bmatrix}\\ u_* \\ l_* \end{bmatrix} = \mathcal{N} \Big(\begin{bmatrix}
\begin{bmatrix}\bm{0}\\\bm{0}\end{bmatrix}
\\0\\0
\end{bmatrix}
, \begin{bmatrix} 
\widehat{K}_{u\r_{\chi}}
 &\begin{matrix}
K_{uu_*}&K_{u l_*}\\
K_{\r_{\chi}u_*}&K_{\r_{\chi}l_*}
\end{matrix}\\
\begin{matrix}
K_{uu_*}^T&K_{\r_{\chi}u_*}^T\\
K_{u l_*}^T&K_{\r_{\chi}l_*}^T
\end{matrix}
&\begin{matrix}
K_{u_*}&K_{u_*l_*}\\
k_{u_*l_*}^T&K_{l_*}
\end{matrix}\end{bmatrix}\Big),
\end{equation}
where 
$$\widehat{K}_{u\r_{\chi}}=\begin{bmatrix}
K_u&K_{u \r_{\chi}}\\
K_{u \r_{\chi}}^T  &K_{\r_{\chi}}
\end{bmatrix}\in \Real^{(n+m)\times (n+m)},$$
$K_u = K_{uu} + \sigma^2_u \bm{I}_u\in \Real^{n\times n}$,
 $ K_{\r_{\chi}} = K_{\r_{\chi}\r_{\chi}}+ \sigma^2_r \bm{I}_{r} \in \Real^{m\times m}$,
  $K_{u_*} = k_{uu}(\x_*,\x_*)$,
$K_{l_*}= k_{ll}(\x_*,\x_*)$
and 
other matrices, e.g.,  $K_{uu_*}, K_{u\r_{\chi}}, K_{u_*l_*}$ 
are defined similarly via the kernels 
$k_{uu}$, $k_{ur}$ and $k_{ul}$.
As mentioned above,  {  $\r_{\chi}=\mathcal{L}u (\chi)$} is actually known as a zero vector since   the differential equation
$\mathcal{L}u=0$ holds everywhere.   So, based on the Bayesian formula 
\begin{equation*}
p(\cdot|\y, \r_{\chi})  = \frac{p(\cdot,\y,\r_{\chi})}{p(\y,\r_{\chi})},
\end{equation*}
and the Gaussian process priors assumption of $p(\cdot,\y,\r_{\chi})$ and $p(\y,\r_{\chi})$, we 
can  get an explicit formula  $p(\cdot|\y, \r_{\chi})$  as below:
 \begin{align}
\label{eq:poster_u}
p(u(\x_*)|\y, \r_{\chi}) &= \mathcal{N}(\bar{u}(\x_*), S_u(\x_*)),\\
p(l(\x_*)|\y, \r_{\chi}) &= \mathcal{N}(\bar{l}(\x_*), S_{l}(\x_*)),
\end{align}
with   
\begin{align}
\bar{u}(\x_*) &= K_{u_*\bullet}\widehat{K}_{u\r_{\chi}}^{-1}\begin{bmatrix}\y\\\r_{\chi}\end{bmatrix},
\label{258}\\
S_u(\x_*)&=K_{uu}(\x_*,\x_*) - K_{u_*\bullet} \widehat{K}_{u\r_{\chi}}^{-1}K_{u_*\bullet}^T,
\label{259}\\
\bar{l}(\x_*) &= K_{l_*\bullet}\widehat{K}_{u\r_{\chi}}^{-1}\begin{bmatrix}\y\\\r_{\chi}\end{bmatrix},
\label{260}\\
 S_l(\x_*)&=K_{ll}(\x_*,\x_*) - K_{l_*\bullet}\widehat{K}_{u\r_{\chi}}^{-1}K_{l_*\bullet}^T,
 \label{261}
\end{align}
where $K_{u_*\bullet} = [K_{u_*u},  K_{u_*\r_{\chi}}]\in \Real^{1\times (n+m)}$ and $K_{l_*\bullet} = [K_{l_*u} ,K_{l_*\r_{\chi}}]\in \Real^{1\times (n+m)}$. 
The posterior variances $S_u(\x_*)$ and $S_l(\x_*)$ can be used as good indicators of how confident the predictions are.  The above results can be easily generalized to 
multiple test points and multiple outputs of different 
derivaties $l$.

Then the estimation of posterior of state $p(u(\x_*)|\y,\r_{\chi})$ and its derivative $p(l(\x_*)|\y,\r_{\chi})$ include the data and differential equation information. Furthermore, such built-in quantification of uncertainty encoded in the posterior variances is a direct consequence of the Bayesian approach adopted in this work. Although not pursued here, this information is very useful in designing a data acquisition plan, often referred to as active learning, which  can be used to optimally enhance our knowledge about the parametric linear equation under consideration.

\section{Initial and Boundary Conditions (IC/BC)}
\label{set:IBC}
The Gaussian process regression method may have a poor prediction near the initial stage or boundary which is caused by imbalance data there. 
To improve the predictions by a given IC/BC,  we employ a{ \it Product of Experts method} which has been widely used  \cite{mayraz2001recognizing,barber2014gaussian,Calderhead2009}, to correct the posteriors of state and its derivatives. 
  We define a normal distribution $p(u | \x, IC/BC)$ which contains the IC/BC information, whose mean is the state value at the point $\x_0$ (the nearest initial or boundary point to $\x$) and whose variance increases  as $\x$ moves away from $\x_0$. The notation $IC/BC$ represents the initial and boundary condition information.
The formula of  Product of Experts  links 
the   statistical models $p(u_*|\x_* , \y, \r_{\chi})$ in \eqref{eq:poster_u} and the normal distribution $p(u_*|\x_*,  IC/BC)$  by 
\begin{equation}
\label{eq:ICBC}
\begin{split}
&p(u_*|\x_*,  \y, \r_{\chi}, IC/BC )\\
& \propto p(u_*| \x_*, \y, \r_{\chi}) \cdot p(u_*| \x_*, IC/BC), 
\end{split}
\end{equation}
where   $p(u_*|\x_*,  \y, \r_{\chi}, IC/BC )$ is the posterior both considering the equation constraint and the IC/BC. We propose the following Gaussian assumption for $p(u_*|\x_*, IC/BC)$
\begin{equation}
\label{eq:icbc}
p(u_*|\x_*, IC/BC) = \mathcal{N}\big(\bar{u}_{IC/BC}(\x_*), S_{IC/BC}(\x_*)\big),
\end{equation}
  where 
$$\bar{u}_{IC/BC}(\x_*) = u(\x_0),\qquad 
S_{IC/BC}(\x_*) = \exp(|| \x_* - \x_0||^2_{\bm{\gamma}_l}) -1.
$$
 Here $\x_0$ is an initial or boundary point closest to $\x_*$ and 
  $\bm{\gamma}_l$ is the same kernel hyper-parameters as in $k_{uu}$. 
  Then the posterior distribution  $p(u_*|\x_*,\y,\r_{\chi},IC/BC)$ in \eqref{eq:ICBC}  is 
  also a Gaussian distribution with the density function   \begin{equation}
\label{eq:posterior_with_IBC}
\begin{split}
p(u_*|\x_*,\y,\r_{\chi},IC/BC) = C_c(\x_*) \cdot p_\mathcal{N}\big(u_*;u_c(\x_*), S_c(\x_*)\big),
\end{split}
\end{equation}
where  $p_{\mathcal{N}}(\cdot\, ; \mu, \Sigma)$ refers to  the probability density function for the Gaussian distribution  
with mean $\mu$ and covariance $\Sigma$, and 
\begin{align*}
C_c(\x_*) &= p_\mathcal{N}(\bar{u} ; \bar{u}_{IC/BC}, (S_u + S_{IC/BC})),\\
u_c(\x_*) &= (S_u^{-1} + S_{IC/BC}^{-1})(S_u^{-1}\bar{u} + S_{IC/BC}^{-1}\bar{u}_{IC/BC}),\\
S_c(\x_*) &= (S_u^{-1} + S_{IC/BC}^{-1})^{-1}.
\end{align*}
$\bar{u}$  is defined by \eqref{258}  and $S_u$ is defined by \eqref{259}.

The formulation of \eqref{eq:ICBC}\eqref{eq:icbc} and \eqref{eq:posterior_with_IBC} can also be extended 
 to the estimation of derivative terms for satisfying the initial and boundary conditions
 if we knew the IC/BC of the derivatives in concern from the given differential equation. 
 
\section{Numerical examples} 
\label{sec:examples}
\subsection{Linear ordinary differential equation}
 In this example $u$ is a function of time and 
 $u, u', u''$ refer to the state, the first order and second order derivative functions respectively. The linear  ODE here is  
\begin{equation}
\label{eq:exam1ode}
u''(t)+bu'(t)+cu(t)=0,
\end{equation}
where $b = 1$ and $c = 3$. The initial condition is given $u(0) = \pi-0.1, u'(0) = 0$, then the   second order derivative  can be computed directly by \eqref{eq:exam1ode},   $ u''(0) = 0.3-3\pi$.
We assume the prior of the sate function $u$ is a zero-mean Gaussian process expressed in \eqref{eq:u_gp}. 
As discussed in Section \ref{set:method}, the equation constraint $r = u'' + b u' + cu$ is also a Gaussian process, $
r\sim \mathcal{GP}(0,k_{rr}(\x,\x';\bm{\gamma})).$
With the property of covariance, the covariance between $r$ and $u$ can be expanded as
\begin{align*}
\Cov(r,u) &=\Cov(u'' + bu' + cu,u)\\
&=\Cov(u'',u) + b\Cov(u',u)+c\Cov(u,u).
\end{align*}
So,  $k_{ru} = k_{u''u} + bk_{u'u}+ck_{u,u}$.
 Similarly, the kernel functions 
 corresponding to covariances $
   \Cov(r,u'), \Cov(r,u'')$ and $\Cov(r,r)$ are expressed as 
 \begin{align*}
 k_{ru'} &= k_{u''u'} + bk_{u'u'}+ck_{u,u'},\\
 k_{ru''} &= k_{u''u''} + bk_{u'u''}+ck_{u,u''},\\
 k_{rr}& = k_{u'' u''} + b^2 k_{u'u'} + c^2  k_{uu} + 2b k_{u''u'}+2c k_{u''u}+ 2bc k_{u'u}.
\end{align*}  

By Section \ref{ssec:predict},
 the posterior distributions $p(u_*|\y, \r_{\chi}) = \N(m_u, \Sigma_u)$, $p(u_*'|\y, \r_{\chi}) = \N(m_{u'}, \Sigma_{u'})$ and $p(u_*''|\y, \r_{\chi}) = \N(m_{u''}, \Sigma_{u''}) $ are  given below
\begin{align*}
m_{u_*} &= K_{\bullet u_*}^T\widehat{K}_{u\r_{\chi}}^{-1}Y,
\quad 
\Sigma_{u_*} =K_{u_*} - K_{\bullet u_*}^T\widehat{K}_{u\r_{\chi}}^{-1}K_{\bullet u_*}\\
m_{u'_*} &= K_{\bullet u_*'}^T\widehat{K}_{u\r_{\chi}}^{-1}Y,
\quad
\Sigma_{u'_*} = K_{u'_*} - K_{\bullet u_*'}^T\widehat{K}_{u\r_{\chi}}^{-1}K_{\bullet u_*'}\\
m_{u''_*} &= K_{\bullet u_*''}^T\widehat{K}_{u\r_{\chi}}^{-1}Y,
\quad
\Sigma_{u''_*} = K_{u''_*} - K_{\bullet u_*''}^T\widehat{K}_{u\r_{\chi}} ^{-1}K_{\bullet u_*''},
\end{align*}
where   $Y = [\y, \r_{\chi}]^T$.
This is  the posterior estimation of state and derivative functions without considering initial condition. Then \eqref{eq:posterior_with_IBC} is applied for the initial conditions for $u,u'$ and $u''$.

\begin{figure}
\centering
\includegraphics[width=0.5\linewidth]{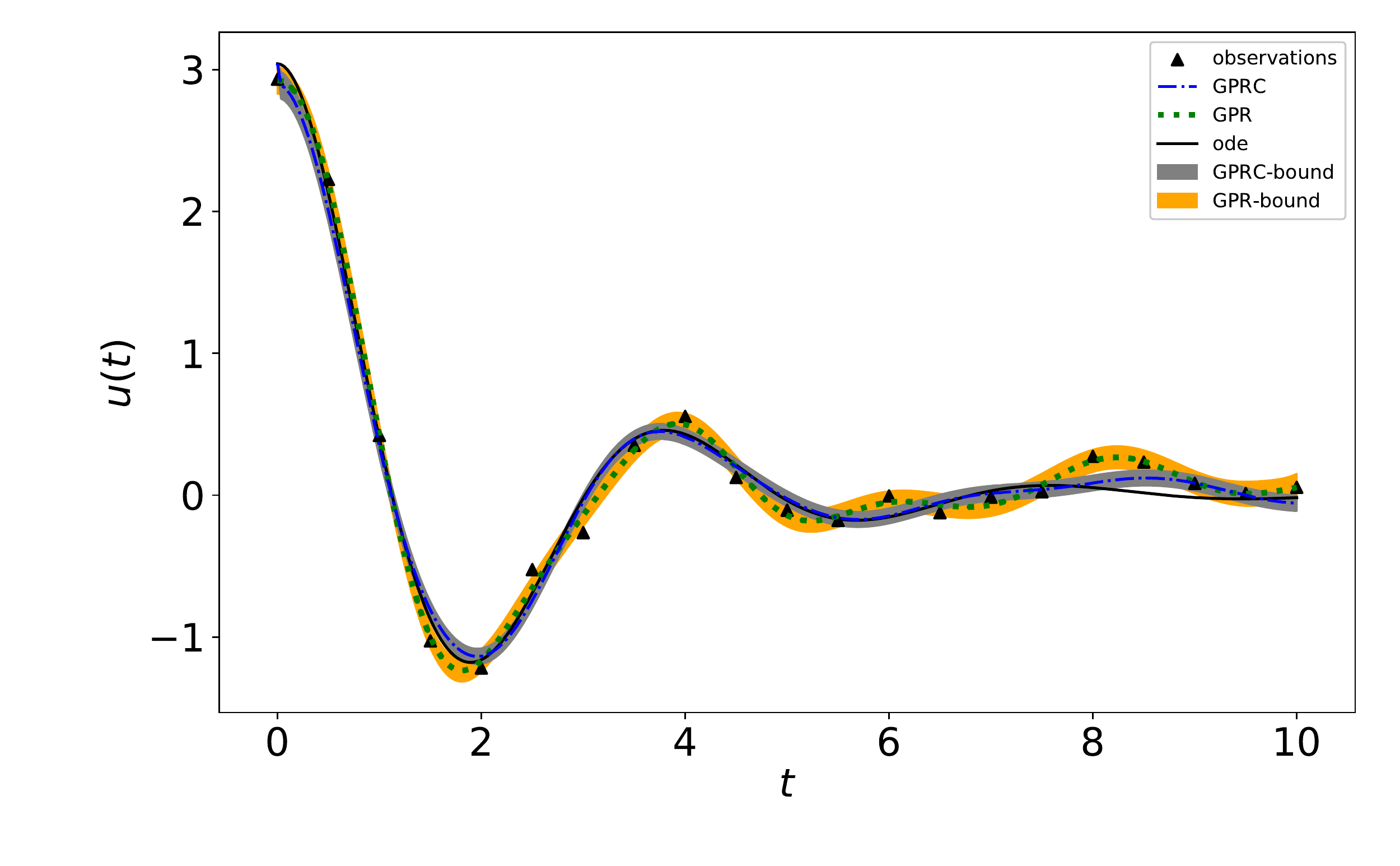}
\caption{The posterior of the state function $u$. The black triangles represent the 
$21$ observation data. The black solid curve represents the
true solution computed by ODE solver. The blue dash-dot/green dotted traces are the inferred posterior means by GPRC and GPR methods, respectively. 
The  confidence bounds  of one    posterior standard deviation are also shown. }
\label{fig:u}                                                                                                                                                                                                                                                               
\end{figure}

\begin{figure}
\centering
\includegraphics[width=0.48\linewidth]{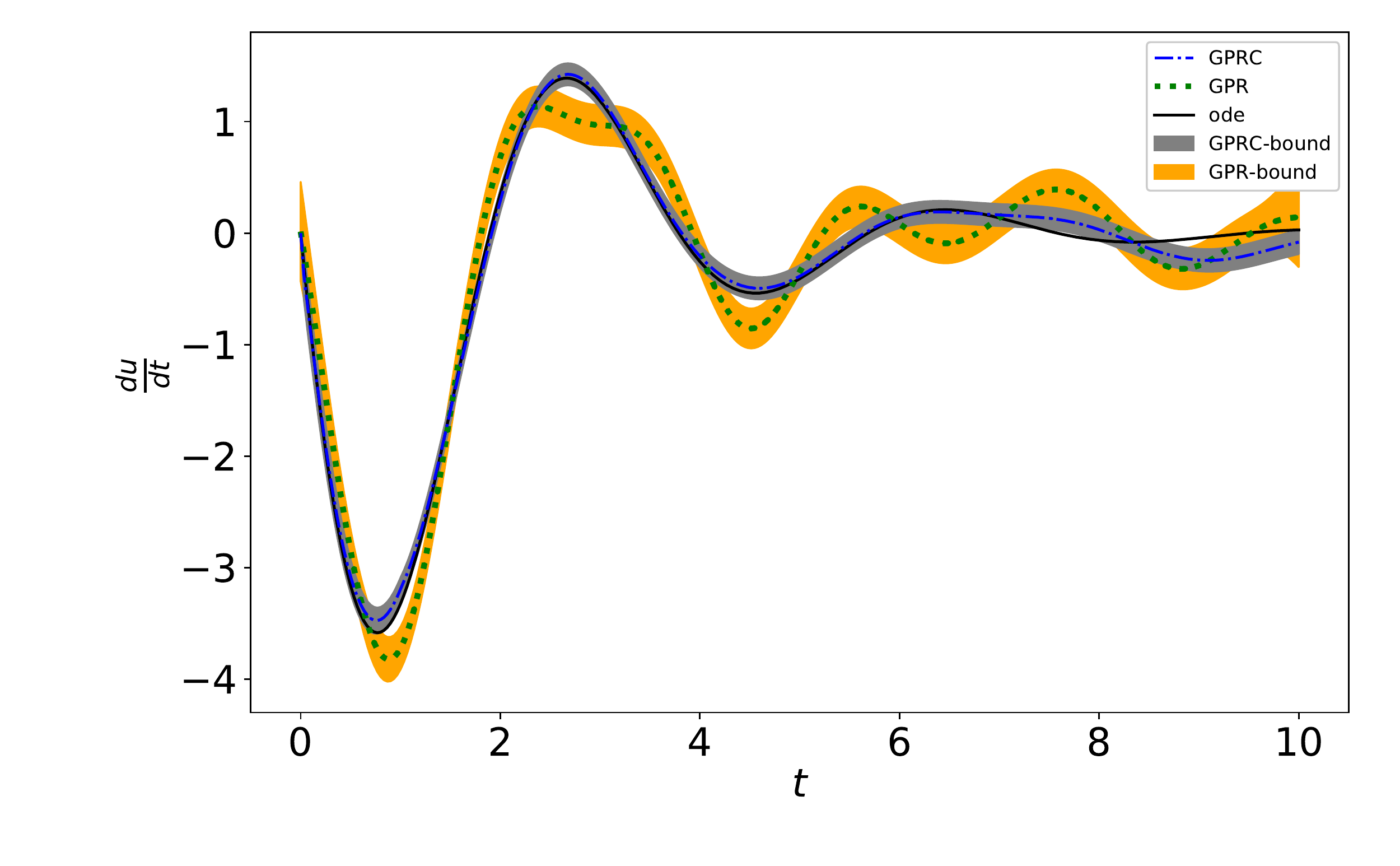}
\includegraphics[width=0.48\linewidth]{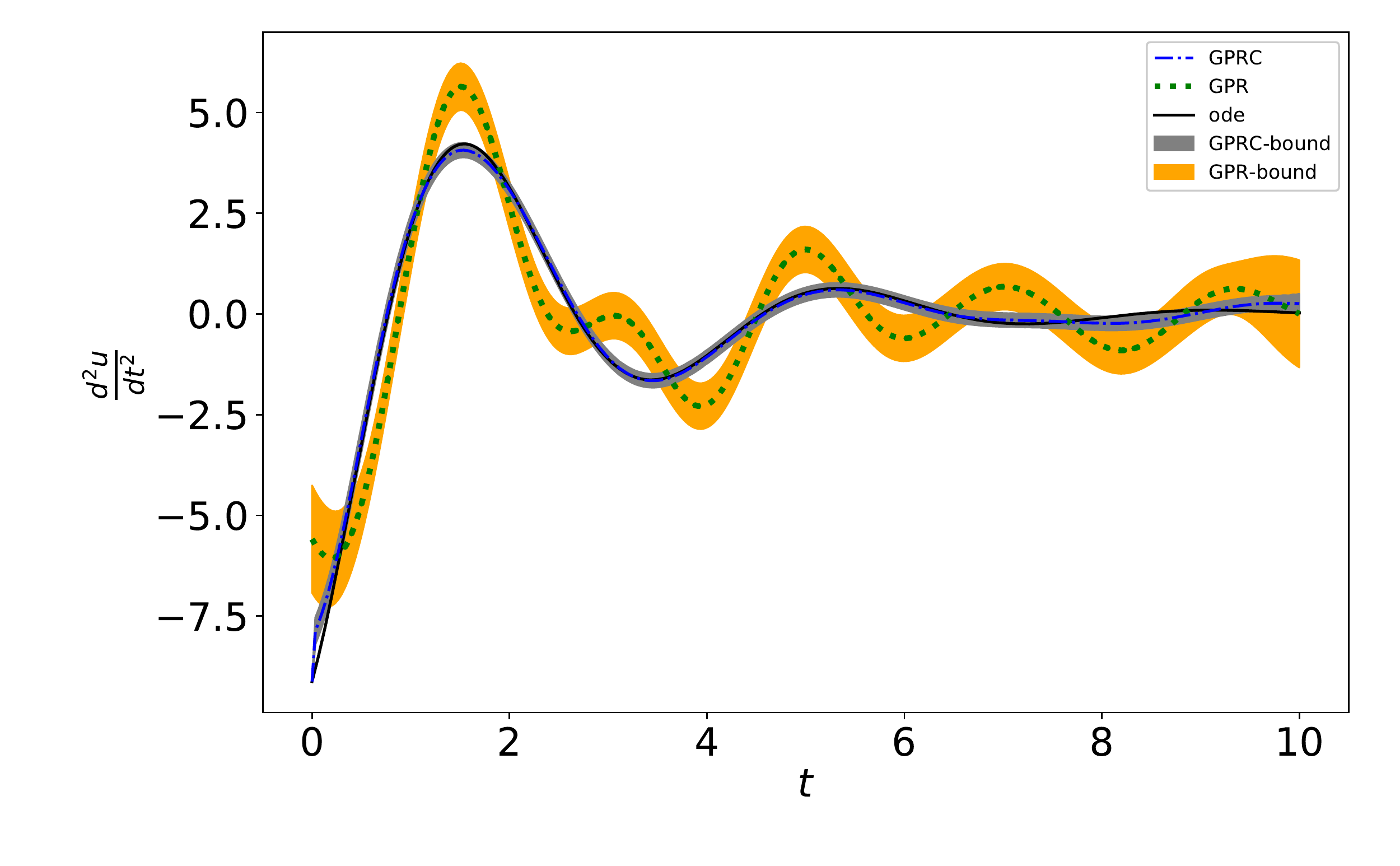}
\caption{The posterior of the first (left) and second (right) order  derivatives. The black 
curves represent  the true solutions. The blue dash-dot/green dotted traces are the inferred posterior means by GPRC and GPR methods, respectively.}
\label{fig:exam2_gh}
\end{figure}

In our experiment, 
there are $21$ observations contaminated by the  additive Gaussian noise with zero-mean and variance $\bar{\sigma}_u^2=0.01$.
We set  the parameter for the equation constraint $\sigma^2_r =0.1$.
In the prediction,  for each $t_*$,  the extended set $\chi$ is chosen as  $60$ points equally spaced in its neighbour $[t_*-3,t_*+3]$.
  Figure \ref{fig:u} and \ref{fig:exam2_gh}  show the posteriors of the state and its derivatives in comparison between GPRC and GPR methods.  The numerical solution from the ODE solver  is regarded as the true solution. 
The results show that the estimations of  all three functions from the GPRC method 
are much closer to the true solution than the traditional GPR and 
demonstrate 
  a great ability to correct the model from the noisy observations with the consideration of equation information. Besides, the GPRC gives a greater extent to the variance reduction of the posteriors estimation since the additional  observations from the equation constraint $\bm r$ \rev{in the training} and $\bm r_{\chi}$ \rev{in the prediction} are  both used.

To examine the effectiveness of our method of incorporating  the equation,
Figure \ref{figure:constraint} shows the residual function   $r(t)=\mathcal{L}{u}(t)$
where  all derivatives in $\mathcal{L}u$  are   the mean function of the estimations from GPRC and GPR.
We see from this figure that the GPRC  with the Product of Experts method in Section \ref{set:IBC}  show an overall smallness for the residuals. 
For a quantitative comparison,  Table \ref{tab:exampe1} presents the {Root Mean Square Error (RMSE) values computed by the posterior mean functions   $u$, $u'$, $u''$ and $r$ with different data noise levels. This confirms the better accuracy of  GPRC method
\rev{with a proper choice of $\sigma_r^2$.}
\rev{We also tested  the effect of  the hyper-parameter  $\sigma_r^2$ in the GPRC method 
by comparing three different values $\sigma_r^2$ in this table.
It is  found that  a  very  large value of $\sigma_r^2$ simply   
reduces  GPRC back to GPR and a  smaller $\sigma_r^2$ 
does not always help reduce the RMSE, which is consistent 
with our interpretation of the regularization effect of $\sigma_r^2$.
Recall  $\sigma_r^2 $  in the matrix $\sigma_r^2\bm{I}_r$ 
 is used  as a     regularization   for the covariance matrix $K_{rr}$ of the residual $r$.
The introduction of this non-zero   factor $\sigma_r^2 $ 
is a beneficial regularization technique. 
Empirically,  we find $\sigma_r^2 \approx 10 \sigma_u^2$ sounds  a good choice
where $\sigma_u^2$ can be first approximated from the traditional GPR.
   }

\begin{figure}[ht]
\vskip 0.2in
\begin{center}
\includegraphics[width=0.5\linewidth]{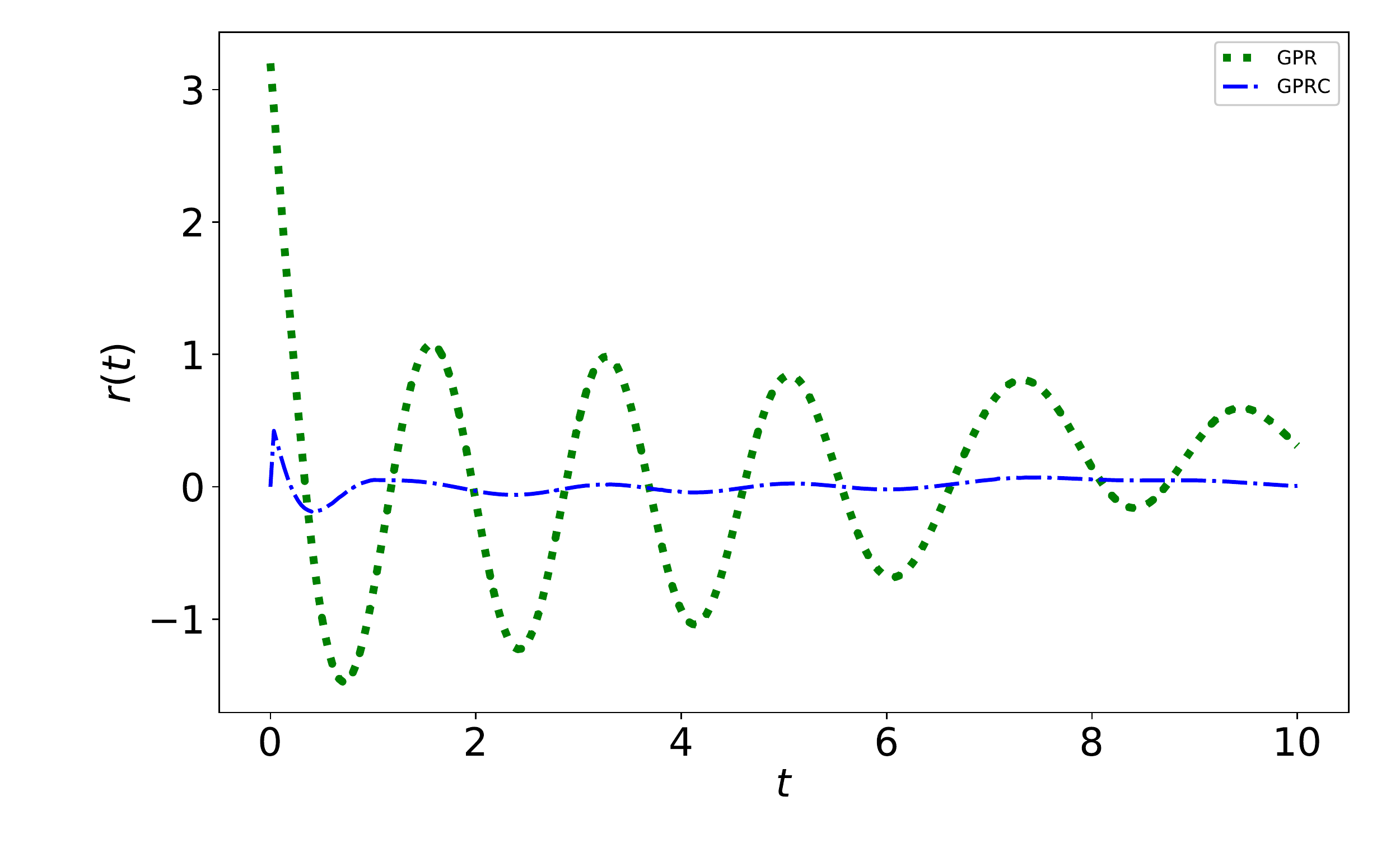}
\caption{The constraint value computed with posterior means   by GPR (green dotted line) and GPRC (blue dash-dot line) method respectively.}
\label{figure:constraint}
\end{center}
\vskip -0.2in
\end{figure}
\begin{table}[!htb]
\caption{RMSE of the posterior mean of state, first derivative, second derivative and residual
 computed by GPR and GPRC when  measurement  noise  with different  variances is injected to the measurement data. $\sigma_r^2$ is the hyper-parameter in GPRC.}
\label{tab:exampe1}
\centering\begin{tabular}{|l|cccc|}
\hline
noise variance: $0.1$ & $u$ & $u'$ & $u''$ & $r$ \\
\hline
GPR     & 0.36& 0.93& 3.32 & 2.09 \\

GPRC($\sigma_r^2=1e{-3}$)  & 0.28& 0.39& 0.75 & 0.69\\

GPRC($\sigma_r^2=1e{-1}$)  & 0.15& 0.20& 0.44 & 0.38\\

GPRC($\sigma_r^2=1e{2}$)  & 0.35& 0.91& 3.20 & 1.93 \\
\hline

 noise variance: $0.05$&  &  &  & \\
\hline
GPR     & 0.16& 0.40& 2.13 & 1.28 \\

GPRC($\sigma_r^2=1e{-3}$)  & 0.20& 0.25& 0.55 & 0.49\\

GPRC($\sigma_r^2=1e{-1}$)  & 0.14& 0.17& 0.41 & 0.36\\

GPRC($\sigma_r^2=5e{-1}$)  & 0.11& 0.16& 0.60 & 0.37\\

GPRC($\sigma_r^2=1e{2}$)   & 0.16& 0.41& 1.89 & 1.12\\
\hline
\end{tabular}
\end{table}

 It is worth noting that too many constraint points is not necessary in prediction process. This viewpoint can be easily verified in this example, shown in Figure \ref{figure:width_step_wrt}. In prediction, we select the constraint points in domain $[t_* - width, t_* + width]$ with spacing $step$, i.e. $t_* = 2$, $width = 1$ and $step = 0.5$, then the constraint points are $\{1, 1.5, 2, 2.5, 3\}$. Figure \ref{figure:width_step_wrt} Left shows the prediction accuracy would tend to constant with the smaller $step$ values. Same tendency occurs in the Figure \ref{figure:width_step_wrt} right, one with respect to the bigger $width$ values. Both figures illustrate one criterion: the prediction ability would not improve with more constraint points, which is an advantage for the computation complexity of GPRC in Eq.\eqref{258}-\eqref{261}.

\begin{figure}[ht]
\vskip 0.2in
\begin{center}
\includegraphics[width=1.0\linewidth]{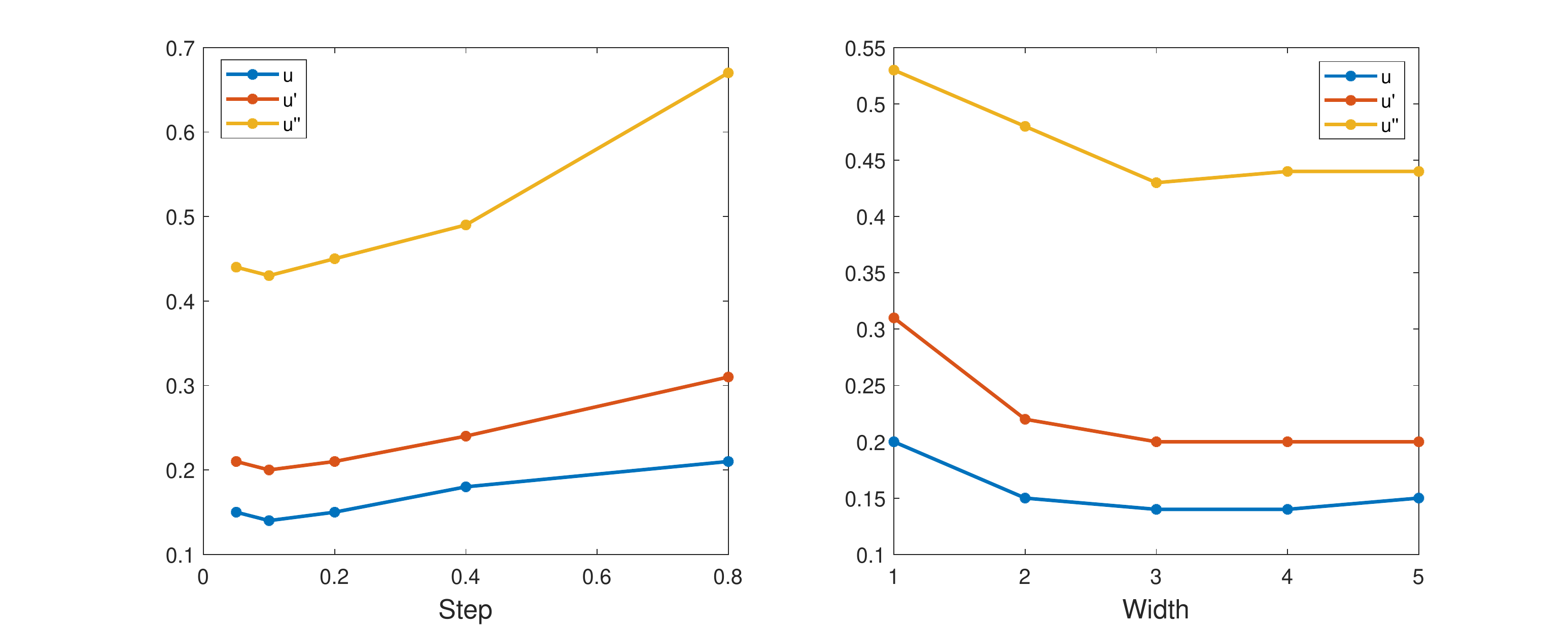}
\caption{RMSE values of $u$, $u'$ and $u''$ with respect to different $step$ and $width$. Left: $width =3$ and Right: $step=0.1$. }
\label{figure:width_step_wrt}
\end{center}
\vskip -0.2in
\end{figure}
 
\subsection{Poisson equation}
Poisson equation describes the spatial variation of a potential function for given source terms and have important applications in electrostatics and fluid dynamics. 
Our setting of Poisson equation is as follows
\begin{equation}
\label{eq:ex2}
\begin{split}
\nabla^2_x u(\x) &= g(\bm{x}), \\
 g(\bm{x}) &= \exp(-x_1)(x_1 -2 +x^3_2 + 6x_2),
 \end{split}
\end{equation}
with Dirichlet  boundary conditions
\begin{equation}
\label{eq:BCs}
\begin{split}
u(0,x_2) &= x^3_2,\\
u(1,x_2) &= (1+x^3_2)\exp(-1),\\
u(x_1,0) &= x_1\exp(-x_1),\\
u(x_1,1) &= (x_1+1)\exp(-x_1),
\end{split}
\end{equation}
and the domain is $ [0,1]\times [0,1]$. The analytic solution is 
\begin{align*}
u(\x) = \exp(-x_1)(x_1+x_2^3).
\end{align*}
This PDE corresponds to the zero value of the residual   $r: = \frac{\partial^2 u}{\partial x_1^2} + \frac{\partial^2 u}{\partial x_2^2}-g $. 
Assume the state is a Gaussian process as \eqref{eq:u_gp}, and then the constraint is also a Gaussian process which is written as  $r(\x)\sim\mathcal{N}(0, k_r)$, where $k_r =  k_{u_1''u_1''}+ k_{u_2''u_2''}+ k_{u_2''u_1''}+ k_{u_1''u_2''}$. The boundary conditions of $u$ and $\frac{\partial^2 u}{\partial x_2^2}$ can be easily obtained by Poisson equation  \eqref{eq:ex2} and Dirichlet BCs \eqref{eq:BCs}. 

\begin{figure}
\begin{center}
\centerline{\includegraphics[width=0.9\linewidth]{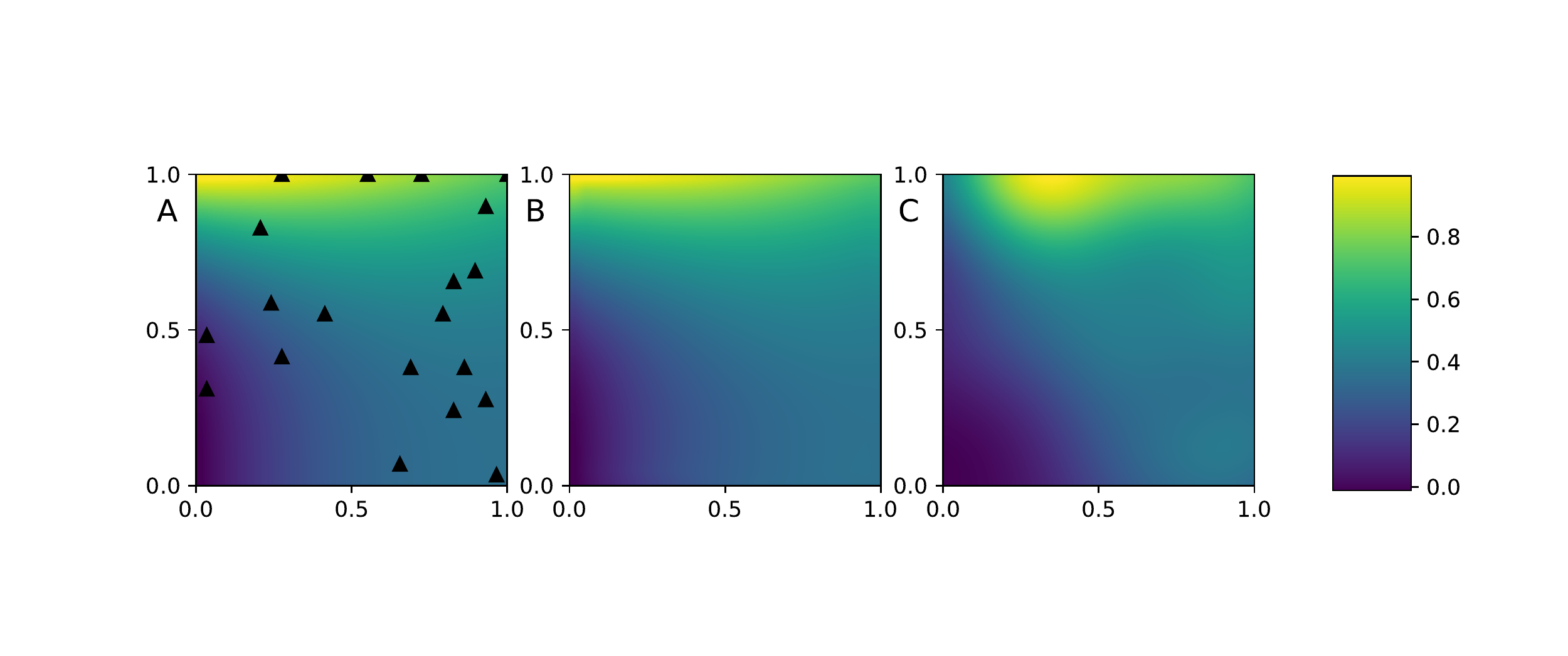}}
\caption{{\textsf A}: The black triangles represent the positions of the observation data and the 
contour plot of the true  solution $u(\x)$. {\textsf B} and {\textsf C} show the posterior mean of state computed by GPRC and GPR, respectively.}
\label{figure:example2_u}
\end{center}
\end{figure}
\begin{figure}
\begin{center}
\centerline{\includegraphics[width=0.9\linewidth]{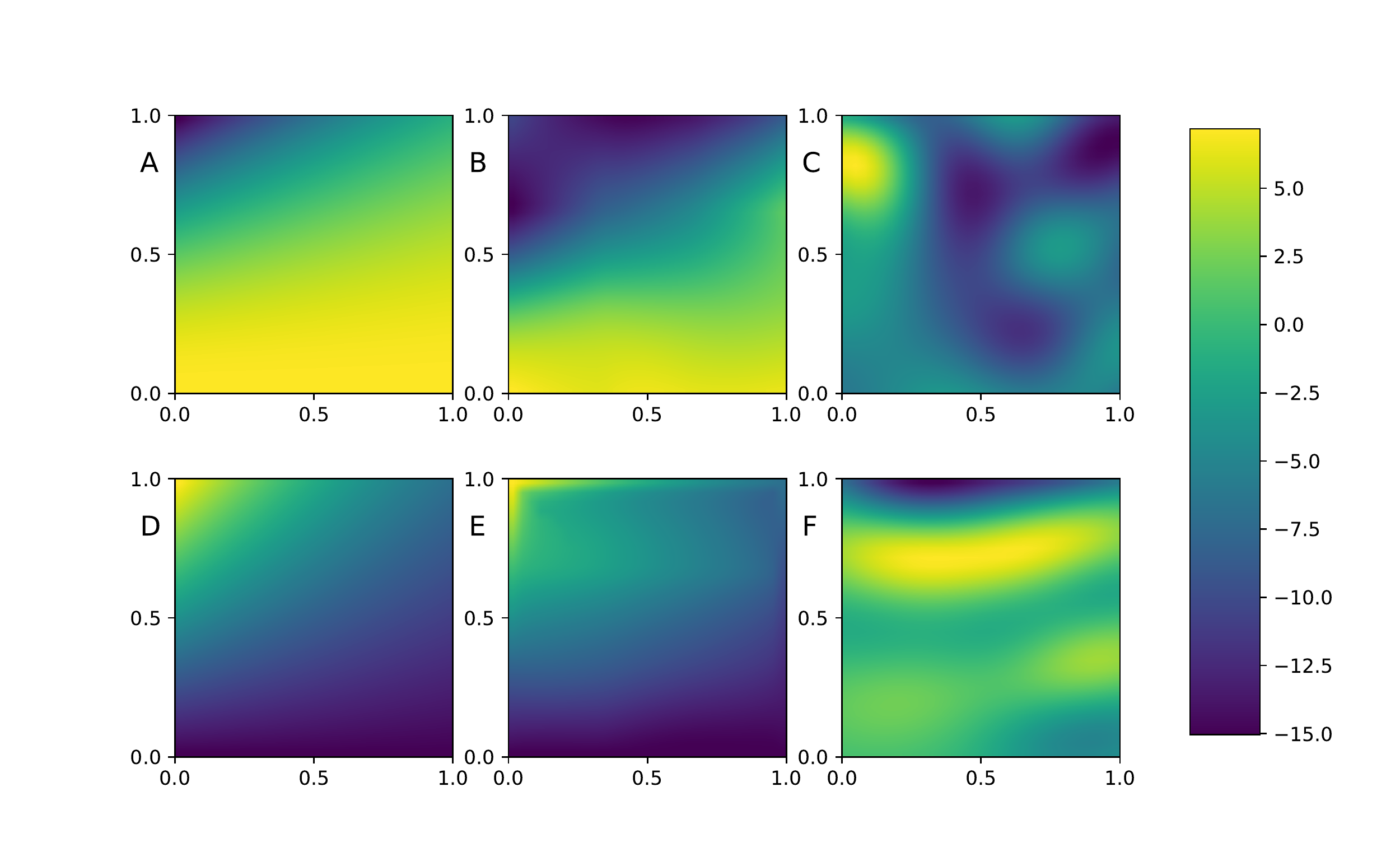}}
\caption{
  The first and second rows represent the estimation of $\frac{\partial^2 u}{\partial x_1\partial x_2}$ and $\frac{\partial^2 u}{\partial x_2^2}$, respectively. {\textsf A} and {\textsf D} show the analytical solutions. {\textsf B} and {\textsf E} represent the posterior means estimated by GPRC. {\textsf C} and {\textsf F} represent the posterior means estimated by GPR.}
\label{figure:example2_h1_h2}
\end{center}
\end{figure}

  $15$ observations of $u$ at $15$ locations in the square domain  were measured as shown in  Figure \ref{figure:example2_u}
with additive noise with   the variance   $\sigma_u^2 =0.01$. 
{  Here we set the slack parameter $\sigma^2_r = 0.3$. The extended set $\chi$ 
in the neighbourhood of a test point $\x_*$ is composed of   $5\times 5$ points equally spaced in square domain $[x_{*1}-0.33, x_{*1}+0.33] \times [x_{*2}-0.33, x_{*2}+0.33]$.} 
{  Figure \ref{figure:example2_u} also shows  the posterior mean functions of the state $u$ estimated by GPRC and GPR, respectively.}
Figure \ref{figure:example2_h1_h2} shows the posterior mean functions of the second order derivatives $\frac{\partial^2 u}{\partial x_1 \partial x_2}$ and $\frac{\partial^2 u}{\partial x_2^2}$, respectively. The predictions of these derivatives 
by GPRC are much  better than the ones by the GPR method. Table \ref{tab:exampe2} shows the RMSE of the posterior mean of state and second derivatives computed by GPR and GPRC respectively. The RMSE values of GPRC is much smaller than the ones computed by GPR, which indicates the advantage of modeling with constraint information.

 \begin{table}[!h]
\centering
\caption{RMSE for the Poisson equation}
\label{tab:exampe2}
\centering\begin{tabular}{l|cccccc}
\hline
     && $u$ && $\frac{\partial^2 u}{\partial x_1 \partial x_2}$   & & $\frac{\partial^2 u}{\partial x_2^2}$   \\
\hline
GPR          & &0.0720  & & 1.91   && 4.71   \\
GPRC         & & 0.0067 && 0.26   && 0.32    \\
\hline
\end{tabular}
\end{table}

 \subsection{Van der Pol equation}\label{VanderPol}
Van der Pol equation is a typical nonlinear ODE which can generate  shock wave solution. It  is defined as 
\begin{equation}
\label{eq:vande}
\frac{d^2u}{dt^2} - \mu(1-u^2)\frac{du}{dt} + u=0,
\end{equation}
where $\mu = 0.5$ and the initial conditions are $u(0) = 2, u'(0)=0$. We can't directly make use of GPRC method to formulate 
the residual of  \eqref{eq:vande} as an aforementioned Gaussian process $r$, since the product of   Gaussian processes is not a Gaussian process anymore.
 Here we propose an iteratively linearization method for nonlinear equations,  motivated by the Picard iteration method\cite{coddington1955theory}. Assume we have an initial guess of the solution $u_0$, then we can rewrite the  equation \eqref{eq:vande} as:
\begin{equation}
\label{eq:vander2}
\frac{d^2u}{dt^2} - \mu(1-u_0^2)\frac{du}{dt} + u=0,
\end{equation}
where the original  $u$ in the nonlinear coefficient  term is replaced by   $u_0$.  
Now   \eqref{eq:vander2} is a linear equation of $u$ and GPRC can be directly applied to estimate the solution $u$ and its derivatives $u', u''$.  The constraint variable for  \eqref{eq:vander2} is then expressed as $r_{0} = \frac{d^2u}{dt^2} - \mu(1-u_0^2)\frac{du}{dt} + u$ and the corresponding value is $0$. This is a quite straightforward linearlization strategy and can be easily applied to most non-linear differential equations. The initial solution guess $u_0$ can be roughly  estimated by the classical GPR (or any other data smoothing method) from observations of $u$.
The new estimation of $u$ from GPRC then serves as a new $u_0$ in 
\eqref{eq:vande} in a recursive way.  \rev{The number of 
such iterations   depends on the quality of learning initial $u_0$ from data,
the specific linearization form of  \eqref{eq:vander2}, 
and the  convergence of Picard iteration. 
So it is  
problem-dependent, but as a practical tool,  
this strategy  is easy to implement 
and in some cases
  it only takes one or two iterations   in practice, as 
shown numerically below in the nonlinear Van der Pol oscillator.
}

\begin{figure}[htpb]
\centering
\includegraphics[width=0.98\linewidth]{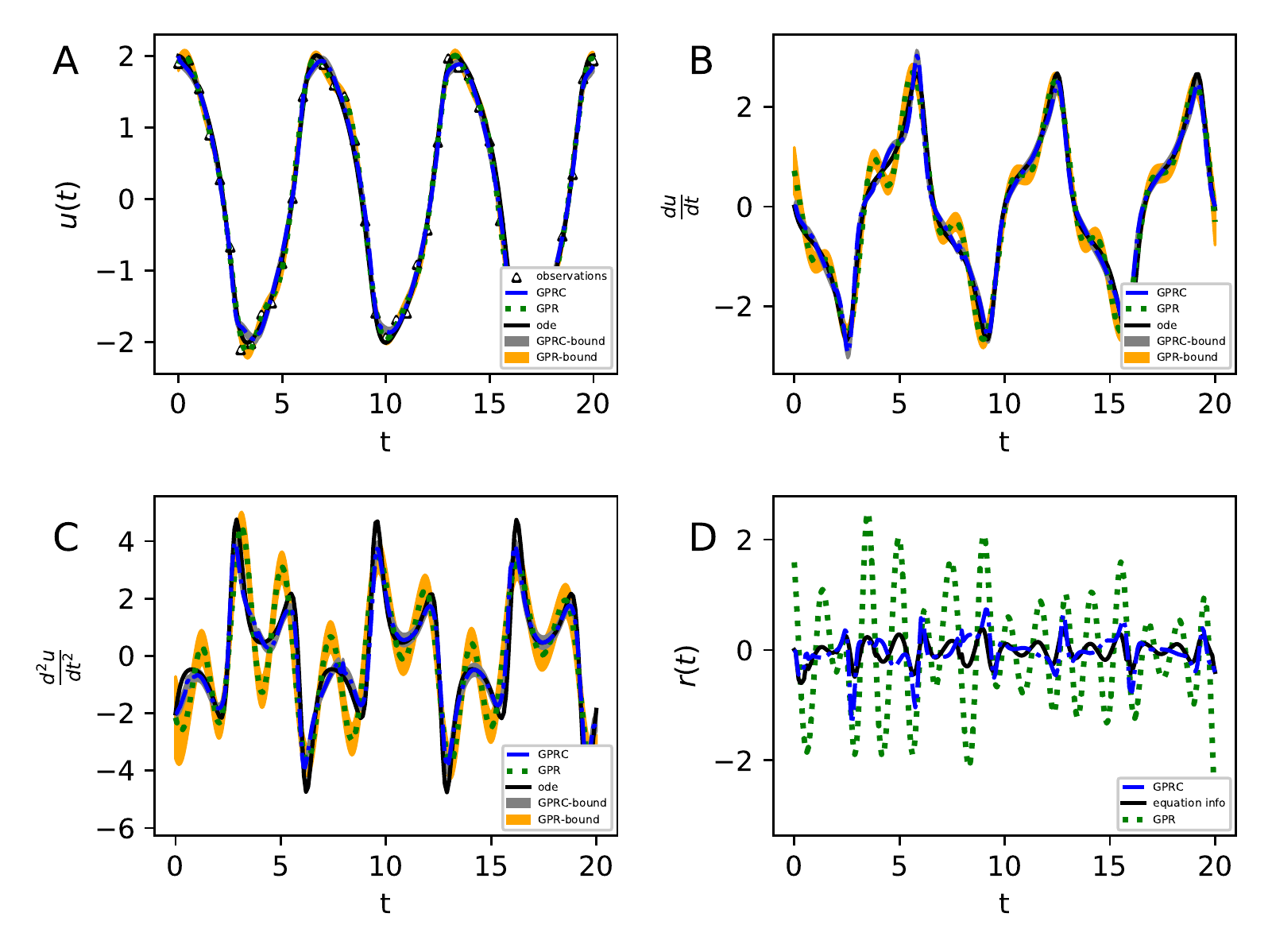}
\caption{ (Van der Pol equation.) {\textsf A, B} and {\textsf C}: The triangles represent the observed data-points of the noisy   process. The black line shows the 
true solution (computed by the ODE solver). The blue dash-dotted annd green dotted lines are, respectively, the inferred posterior means by GPRC and GPR method with 
a confidence band of $\pm$ one posterior standard deviation. {\textsf D}: The black line shows the residual $r_{0}$ of the  linearlization equation  \eqref{eq:vander2}. The blue dash-dotted
and the green dotted lines are  the true (nonlinear residual)  in \eqref{eq:vande} computed on the posterior means  respectively by GPRC and GPR.}
\label{fig:predict_ughr}
\end{figure}

40 observations are measured with  equal space in the time interval $[0, 20]$, contaminated by the additive Gaussian noise with variance   $\sigma^2_u =0.01$. 
  The extend set $\chi$ for each test point $x_*$ is chosen as $4$ points equally spaced in $[x_*-0.2,x_*+0.2]$ and the hyper-parameter $\sigma^2_r = 0.1$. 
  The initial $u_{0}$ is from GPR with the given $40$ observations. 
  Here we only take {\it one} iteration of Picard approximation to obtain a very good result  and  we found more iterations  did not improve the posterior estimation further.
Figure \ref{fig:predict_ughr} compares  the result  from GPR and GPRC.  Although both  methods perform well on the estimation of the state variable $u$,  for the first order derivative function, the GPR method starts to show spurious small oscillations
and for the   second order derivative, the GPR method produces the erratic phase and amplitude.
By contrast, the GPRC method gives a consistent and robust   posterior prediction for the derivatives from zero order to second order.
Note that GPRC here did not solve the Van del Pol equation on the whole interval $[0,20]$ and the extended set  $\chi$ only has a width $0.6$.

Panel \textsf{D} of Figure \ref{fig:predict_ughr}  further  shows the residual $r_{0}$ of the linearized equation (black line) and the true $r$ of the Van der Pol equation (blue line),
which confirms that just one Picard iteration  in this example has achieved quite good results. 
 The RMSE (Root Mean Square Error) of  the nonlinear constraint $r$  with increased  measurement  noise variances are summarized in  Table \ref{tab:exampe3} and  illustrates the trend of convergence of the GPRC algorithm with only one Picard iteration.

  \begin{table}[!htb]
\caption{(Van der Pol equation) The RMSE of the true (nonlinear) residual $r$ based on  the mean function from GPRC with observation contaminated by different noise variances.
}
\label{tab:exampe3}
\centering\begin{tabular}{l|cccccc}
\hline
noise variance     && 0.10    && 0.05    && 0.01 \\
\hline
RMSE          && 0.0128   && 0.0094    && 0.0079\\
\hline
\end{tabular}
\end{table}

\rev{
\subsection{Application to identify the parameter $\mu$}
}
\label{ssec:ip}

\rev{ We continue to consider the Van del Pol model
and explain  how the   robust estimation of derivatives 
by GRPC 
can help improve the method for the problem of parameter identification $\mu$ .
Our previous  numerical results are all associated with a given parameter $\mu=0.5$ in \eqref{eq:vande}.
  We denote  this  ground truth as $\mu^{\star}=0.5$ and make $\mu$ a generic variable.
Let $(t_{i},y_{i})$ be  $n$ given observations  of $u(t)$ on the given location $t_{i}$. The general   methodology \cite{xun2013parameter,brunton2016,rudy2017data}
is to estimate  a function $\widehat{u}(t)$ first
from the data, for instance, by minimizing the sum-of-squared loss $\sum_{i}|y_{i}-\widehat{u}_{i}|^{2}$ where $\widehat{u}_{i}=\widehat{u}(t_i)$
and then to find the optimal parameter $\mu$ by minimizing the squared error of the equation's residual $\sum_{j} |r_{j}|^{2}$ on a set of $m$ design points $\{\widehat{t}_{j}\}$ (which can be either the same as the original measurement
locations or a new set with $m>  n$).
We can write the sum of two losses as follows (with the equal weight for both losses for simplicity): 
\begin{equation}
\label{eq:loss}
L_{loss}(\mu) = \frac{1}{n}\sum_{i=1}^n (y_i - \widehat{u}_i)^2 + \frac{1}{m}\sum_{j=1}^{m} \bigg(\widehat{\frac{d^2u_j}{dt^2}} - \mu(1-\widehat{u}_j^2)\widehat{\frac{du_j}{dt} }+ \widehat{u}_j\bigg)^2.
\end{equation}
The key issue here in our focus  is  the computation of derivatives in the second part:  $\widehat{\frac{du_j}{dt}}$, $\widehat{\frac{d^{2}u_j}{dt^{2}}}$ are just  $\frac{d }{dt} \widehat{u}(t)$ 
and $\frac{d^{2} }{dt^{2}} \widehat{u}(t)$   at $t=\widehat{t}_j$  or estimated alternatively. The traditional method, such as GPR, estimates $\widehat{u}(t)$ only based on the data $(t_i,y_i)$, 
and the derivatives  in \eqref{eq:loss} are the numerical or analytical 
differentials of $\widehat{u}$. So   the obtained function $\widehat{u}$ is independent of $\mu$ and   \eqref{eq:loss}
is exactly a quadratic function of $\mu$ for this example (the first term has no effect in identifying  $\mu$).
However, when our GPRC is applied to this problem, for {\it each} $\mu$,  we use the data $(t_{i},y_{i})$ and the equation associated with this postulated $\mu$ to 
jointly estimate $\widehat{u}(t)$, $\widehat{\frac{du}{dt} }(t)$ 
 $\widehat{\frac{d^2u}{dt^2}}(t) $ to compute  $L_{loss}(\mu)$   
 and consequently,   all  hatted  terms in  \eqref{eq:loss}
 $L_{loss}(\mu)$ depend  on $\mu$ and  $L_{loss}(\mu)$ no longer has a simple 
quadratic expression.
The optimal $\mu$ is then the minimizer of this $L_{loss}(\mu)$.
In summary,   the application of GPRC 
for parameter identification problem, conceivably,  is to solve $\displaystyle \min_{\mu}\min_{\widehat{u},\widehat{u'}, \widehat{u''}} L_{loss}(\mu; \{t_i, y_i\})$ where the $\min$ inside is  the GPRC.
 }

\rev{  
 To  demonstrate the   importance of getting the right estimation of derivatives used in the loss function $L_{loss}$,  Figure \ref{fig:mu} shows the values of $L_{loss}$ 
    with respect to the parameter $\mu$. Note the ground truth $\mu^{\star}=0.5$. 
    With the  same   $n=40$ observations,  the minimizer in the first subplot corresponding to GPRC  can recover this parameter between $0.4\sim0.5$, 
    while the   minimizer in the middle subplot corresponding to GPR gives  a completely wrong answer
    because the derivatives used are not as accurate and robust  as from GPRC  even at the true $\mu^\star$. 
 If we are able to use a large  number of observations of $u$, say $n=400$ in the right subplot of  Figure \ref{fig:mu},
 the traditional GPR is then capable to find the correct optimal parameter $\mu^{\star}$.
 So,  the traditional method like  GPR method to estimate the function $\widehat{u}(t)$ only 
 and to rely on numerical/analytical differentiation of $\widehat{u}(t)$ works {\it only} if the 
number of observations is sufficiently large. The benefit of the new GPRC method advocated here
has a potential advantage when the available data points are limited.}

\begin{center}
  \begin{figure}[htpb]
\includegraphics[width=1.0\linewidth]{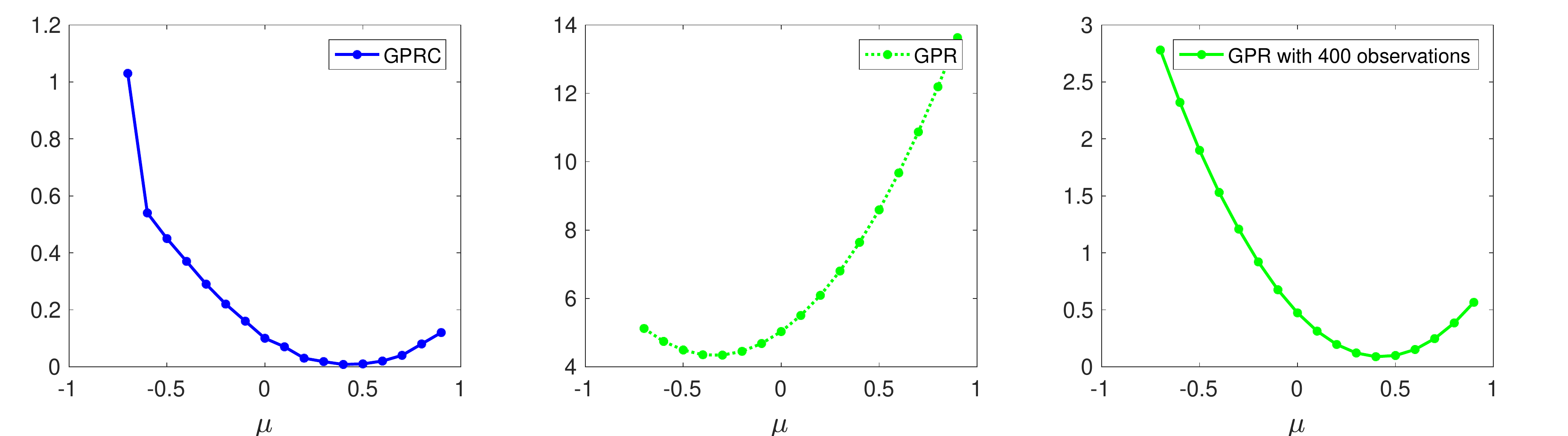}\caption{  {\it Left and Middle}:   $L_{loss}$ in \eqref{eq:loss} with respect to the parameter $\mu$, computed by GPRC (left) and GPR (middle), respectively, based on the same measurements on 40 locations in Section 5.3.  {\it Right}:   $L_{loss}$ with respect to the parameter $\mu$, computed by GPR from
a  larger dataset of  $400$ observations. }
\label{fig:mu}
\end{figure}
\end{center}

 \section{Conclusions}\label{sec:conclusions}

In this work, 
we have shown how to  improve  the  accuracy and robustness for  the numerical estimation 
of  derivatives from the noisy state data by 
our  new method of  Gaussian process regression with constraint (GPRC)  for  linear  and nonlinear differential equations.
Explicit posteriors with uncertainty information are obtained in the Bayesian framework
for the joint multi-dimensional GPR. For nonlinear differential equations,  a  strategy of   linearization method   motivated from the Picard iteration is applied.  
From the perspective of   incorporating the differential equations into the Gaussian process regression (GPR), 
our work is a kind of  physics-informed Gaussian process regression,
compatible with  the recent awareness of  the importance of  combining  the observations of solution data and the underlying  physical model \cite{raissi2019physics}.
 
\rev{
 An important toy example of Van de Pol equation  has been presented 
to show, when applied  to the parameter identification problem,  the joint estimation of the solution and its derivatives by our new method
make an important contribution   to identifying the    missing parameter correctly 
with fewer samples.
The method developed here may have a potential application for the more complication problems like parameter identification \cite{rudy2017data}.
It is  foreseen that the price 
to pay could  be  the extra optimization costs  
 since  all terms in $L_{loss}(\mu)$ in \eqref{eq:loss} 
involve the unknown parameter $\mu$.  It would 
be interesting to  develop certain fast sequential  methods  to improve the efficiency for   such problems.}

So far, the equations we have considered  are deterministic with   known
or missing  coefficients, and the hyper-parameter parameter $\sigma_r^2$ is beneficial in practice as regularization
 to allow some prior distribution for the zero residual.
For the stochastic differential equations with random coefficients, 
one may include this uncertainty of residual into the likelihood function in the same Bayesian framework; 
however with the restriction of Gaussian assumption,  the approach of GPR may not be applicable except in some special cases in \cite{raissi2017machine}.












\bibliographystyle{plain}

\bibliography{muban}
\end{document}